ARTICLE TYPE

Research letter

TITLE

Expiratory variability index (EVI) is associated with asthma risk, wheeze and lung function in infants with recurrent respiratory symptoms

AUTHORS AND INSTITUTIONS

Seppä, Ville-Pekka[1]; Gracia Tabuenca, Javier[2]; Kotaniemi-Syrjänen, Anne[3]; Malmström, Kristiina[3]; Hult, Anton[1]; Pelkonen, Anna S[3]; Mäkelä, Mika J[3]; Viik, Jari[2]; Malmberg, L Pekka[3]

1: Revenio Research Ltd., Vantaa, Finland
2: Tampere University, Faculty of Medicine and Health Technology, Tampere, Finland
3: Helsinki University Central Hospital, Department of Allergy, Helsinki, Finland

Corresponding Author:

Seppä, Ville-Pekka
Postal address: Revenio Research Ltd., Äyritie 22, FI-01510 Vantaa, Finland
E-mail: vps@iki.fi
Telephone number: +358 445062797

ABSTRACT

Recurrent respiratory symptoms are common in infants but the paucity of lung function tests suitable for routine use in infants is a widely acknowledged clinical problem. In this study we evaluated tidal breathing variability (expiratory variability index, EVI) measured at home during sleep using impedance pneumography (IP) as a marker of lower airway obstruction in 36 infants (mean age 12.8 [range 6-23] months) with recurrent respiratory symptoms. Lowered EVI was associated with lower lung function (VmaxFRC), higher asthma risk, and obstructive symptoms, but not with nasal congestion. EVI measured using IP is a potential technique for lung function testing in infants.

Analysis of tidal breathing flow-volume (TBFV) curves is a convenient method to assess lung function in infants. We have shown that impedance pneumography (IP) is a valid method to assess overnight TBFV profiles in young children [1] and infants [2] and that certain specific curve shapes are related to symptom presentation and asthma risk [3]. We have also shown that the reduced variability of TBFV curves may be an indicator of airway obstruction in children [4–6]. However, in infants, it is unknown whether overnight TBFV variability is associated with respiratory symptoms or asthma risk.

TBFV variability has been expressed in several ways [4, 5, 7–10]. The recently introduced expiratory variability index (EVI) is derived by calculating correlations between expiratory TBFV curves, with low EVI indicating reduced variability of the curve shapes [6].

In this study, we hypothesised that TBFV variability, expressed as EVI, would be associated with wheeze and asthma risk in infants with recurrent respiratory symptoms.

The patient sample was described in detail earlier [3]. Briefly, between January 2014 and May 2017 we recruited 43 infants referred to the Department of Allergy, Helsinki University Hospital, for infant lung function testing because of recurrent respiratory symptoms such as wheeze, cough, and/or laborious breathing.

Children were classified by asthma risk by using loose criteria of the modified Asthma Predictive Index (mAPI): Children with a history of multiple episodes of wheeze (at least 2) and who fulfilled one of the major criteria (parental history of asthma, atopic dermatitis, sensitization to respiratory allergens), were considered to have a high risk of asthma. Children who did not have a history of wheeze or did not fulfil any of the major criteria above were considered to have low risk. Infants who fell between the criteria of high and low risk were considered to have an intermediate risk. Children with current respiratory infection were excluded.

Lung function was determined by measuring the maximal flow at functional residual capacity (VmaxFRC) using rapid thoracic compression (Babybody Masterscreen, Jaeger Ltd., Germany) under sedation, and expressed as z scores, adjusting for weight, length, and/or sex [11]. Fractional exhaled nitric oxide (FeNO) was measured with the tidal online technique [12]. In a subgroup of infants (n=14), airway hyperresponsiveness (AHR) to methacholine was assessed as described earlier [13] and expressed as the provocative dose causing a reduction of 40% in VmaxFRC (PD40VmaxFRC). After lung function tests, an IP device (university-developed prototype [14]) was installed on the subjects for overnight recording at home. The study was approved by the institutional paediatric ethics committee of Helsinki University Central Hospital (approval no. 53/13/03/03/2012). All parents gave written informed consent.

The previously recorded raw impedance signals were re-processed using commercial software (Ventica Analytics 2.1.0, Revenio Research Ltd., Finland) for the expiratory variability index (EVI).

EVI is derived by calculating Pearson correlations between all partial (15-45 % of exhaled volume [5]), averaged (5-minute window) TBFV curves recorded from the duration of the night's sleep resulting in a large number of correlation values. The output is given as EVI=[log10[IQR(r)]+2]*10, where IQR(r) is the inter-quartile range of the correlations. EVI thus measures the variability (dissimilarity) between TBFV curves compared to each other from all parts of the night, low EVI indicating low variability.

Parametric tests (Student's t test, ANOVA linear trend test) were used for comparing EVI between categories if EVI was normally distributed in each category (Shapiro-Wilks test p value above 0.05). Otherwise non-parametric test (Mann-Whitney U test) was used (only for nasal congestion and atopic eczema). Association of EVI with continuous variables (VmaxFRC, PD40VmaxFRC, FeNO) was assessed by Pearson correlation. No adjustments for multiple comparisons were made.

Overnight IP recordings at home were attempted on 43 infants of which 7 were rejected by the processing software because of bad electrode contact or malfunctioning of the prototype recorders. In 36 children IP measurements were successful allowing the calculation of EVI result. These children were 12.8 (5.1, 6.0-23.0) (mean, SD, range) months old, 20 (56 %) of them had atopic eczema, 9 (25 %) positive skin prick test finding and they had had 1.7 (1.5) past wheeze episodes.

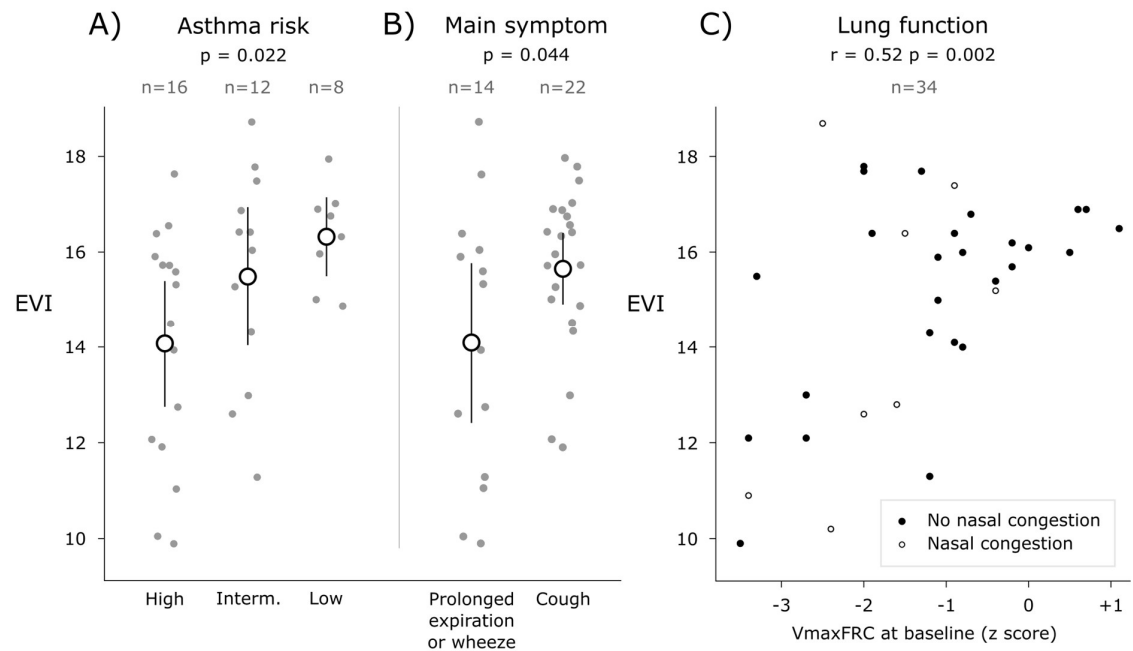

*Figure 1. Expiratory variability index (EVI) compared with A) asthma risk groups, B) main symptom, and C) lung function. Circles and whiskers denote, respectively, mean, and its 95% confidence interval. Dots represent results of individual patients. Two patients did not have VmaxFRC results available.*

EVI was different between asthma risk categories (ANOVA linear trend test p=0.022, figure 1A) being lowest in the high asthma risk children. Children with prolonged expiration (n=14) or wheeze as the main symptom had lower EVI than those with only cough (n=22) (Student's t test p=0.044, figure 1B). EVI showed positive correlation with VmaxFRC (Pearson correlation r=0.52, p=0.002, figure 1C).

EVI values from children with nasal congestion (n=10) were not statistically different from those without congestion (Mann-Whitney U test p=0.50, figure 1C). This is in line with a recent finding in older children [6]. EVI was not correlated with age (Pearson correlation r=-0.06).

In overnight recordings, EVI was not significantly associated with AHR (Pearson correlation r=0.11, p=0.70), nor with atopic eczema (Mann-Whitney U test p=0.40), prick test (Student's t test p=0.49) or FeNO (Pearson correlation r=0.21, p=0.24).

In our previous study, we demonstrated that certain features of the shape of the expiratory flow-volume curve are associated with asthma risk and lung function in infants with respiratory symptoms [3]. In this study, we focused instead only on the variability of the curve shapes, expressed as EVI. Although the within-subject variation was large, we found EVI to associate with asthma risk, main symptom and lung function. The associations were only modest which may be attributed to small sample size and the low specificity and heterogeneity of the outcomes for asthma.

In this observational study, the results indicate reduced TBFV variability in infants with wheeze, which agrees with studies in in adults with asthma [7] and COPD [8], children with asthma [4, 9] and infants with bronchopulmonary dysplasia [10]. This reduced variation in TBFV pattern which has been attributed to reduced adaptability of the respiratory system during mechanical loading by airway obstruction [7] should be distinguished from the other phenomenon of increased variation in respiratory impedance or diurnal peak flow commonly known as characteristics of asthma [15].

EVI did not show association with AHR to methacholine. We speculate that decreased TBFV variability as expressed by EVI, is more closely related to reduced lung function and adaptive mechanisms of tidal breathing than to AHR [3]. VmaxFRC as a lung function measure has a wide intersubject variability in infants which explains the modest association with EVI. Moreover, the mechanisms of reduced VmaxFRC vary between infants with respiratory symptoms. Depending on the mechanical resistive or elastic loading, neural and mechanical adaptive changes in breathing pattern may change and affect EVI differently [16]. It has been shown that sleep stages are a major contributor to normal TBFV curve changes which constitute normal (high) EVI [17].

Cough as the main symptom may have heterogeneous aetiology and it is an unsure sign of airway obstruction. This agrees with our finding of reduced EVI being more common in infants with wheeze as the main symptom. We acknowledge that due to the lack of a healthy control group and longitudinal data we are unable to estimate the proportion of abnormal EVI results in different patient groups or assess the specificity of EVI to estimate asthma risk in infants with recurrent respiratory symptoms. Furthermore, in this study, we did not estimate the interactions between sleep architecture and EVI, which need further studies.

In conclusion, the results suggest that variability of nocturnal tidal breathing, as quantified by the EVI parameter measured using the IP technique, is associated with asthma risk, symptoms and lung function in infants with recurrent respiratory symptoms. Provided that the findings are repeated in larger cohorts of infants, this technique may prove to be a useful, easy tool for objective assessment of respiratory symptoms in this age group where lung function testing is difficult.

ACKNOWLEDGMENTS

We acknowledge Professor Seppo Sarna, Helsinki University Hospital, Helsinki, Finland for advices regarding statistical reporting.

REFERENCES

1. Seppä V-P, Pelkonen AS, Kotaniemi-Syrjänen A, Mäkelä MJ, Viik J, Malmberg LP. Tidal breathing flow measurement in awake young children by using impedance pneumography. J Appl Physiol 2013; 115: 1725–1731.

2. Malmberg LP, Seppä V-P, Kotaniemi-Syrjänen A, Malmström K, Kajosaari M, Pelkonen AS, Viik J, Mäkelä MJ. Measurement of tidal breathing flows in infants using impedance pneumography. Eur Respir J 2017; 49: 1600926.

3. Gracia-Tabuenca J, Seppä V-P, Jauhiainen M, Kotaniemi-Syrjänen A, Malmström K, Pelkonen A, Mäkelä MJ, Viik J, Malmberg LP. Tidal breathing flow volume profiles during sleep in wheezing infants measured by impedance pneumography. Journal of Applied Physiology 2019; 126: 1409–1418.

4. Seppä V-P, Pelkonen AS, Kotaniemi-Syrjänen A, Viik J, Mäkelä MJ, Malmberg LP. Tidal flow variability measured by impedance pneumography relates to childhood asthma risk. European Respiratory Journal 2016; 47: 1687–1696.

5. Seppä V-P, Hult A, Gracia-Tabuenca J, Paassilta M, Viik J, Plavec D, Karjalainen J. Airway obstruction is associated with reduced variability in specific parts of the tidal breathing flow–volume curve in young children. ERJ Open Research 2019; 5: 00028–02019.

6. Seppä V-P, Paassilta M, Kivistö J, Hult A, Viik J, Gracia-Tabuenca J, Karjalainen J. Reduced expiratory variability index (EVI) is associated with controller medication withdrawal and symptoms in wheezy children aged 1-5 years. Pediatric Allergy and Immunology [Internet] 2020; Available from: https://doi.org/10.1111/pai.13234.

7. Veiga J, Lopes AJ, Jansen JM, Melo PL. Airflow pattern complexity and airway obstruction in asthma. J Appl Physiol 2011; 111: 412–419.

8. Dames KK, Lopes AJ, de Melo PL. Airflow pattern complexity during resting breathing in patients with COPD: Effect of airway obstruction. Respiratory Physiology & Neurobiology 2014; 192: 39–47.

9. Hmeidi H, Motamedi-Fakhr S, Chadwick EK, Gilchrist FJ, Lenney W, Iles R, Wilson RC, Alexander J. Tidal breathing parameters measured by structured light plethysmography in children aged 2–12 years recovering from acute asthma/wheeze compared with healthy children. Physiological Reports 2018; 6: e13752.

10. Usemann J, Suter A, Zannin E, Proietti E, Fouzas S, Schulzke S, Latzin P, Frey U, Fuchs O, Korten I, Anagnostopoulou P, Gorlanova O, Frey U, Latzin P, Proietti E, Usemann J. Variability of Tidal Breathing Parameters in Preterm Infants and Associations with Respiratory Morbidity during Infancy: A Cohort Study. The Journal of Pediatrics 2018; .

11. Hoo A-F, Dezateux C, Hanrahan JP, Cole TJ, Tepper RS, Stocks J. Sex-Specific Prediction Equations for V˙maxFRC in Infancy. Am J Respir Crit Care Med 2002; 165: 1084–1092.

12. Hall GL, Reinmann B, Wildhaber JH, Frey U. Tidal exhaled nitric oxide in healthy, unsedated newborn infants with prenatal tobacco exposure. J. Appl. Physiol. 2002; 92: 59–66.

13. Kotaniemi-Syrjänen A, Malmberg LP, Pelkonen AS, Malmström K, Mäkelä MJ. Airway responsiveness: associated features in infants with recurrent respiratory symptoms. Eur. Respir. J. 2007; 30: 1150–1157.

14. Vuorela T, Seppä V-P, Vanhala J, Hyttinen J. Design and implementation of a portable long-term physiological signal recorder. IEEE Transactions on Information Technology in Biomedicine 2010; 14: 718–725.

15. Frey U, Brodbeck T, Majumdar A, Taylor DR, Town GI, Silverman M, Suki B. Risk of severe asthma episodes predicted from fluctuation analysis of airway function. Nature 2005; 438: 667–670.

16. Carroll JL, Donnelly DF. Respiratory physiology and pathophysiology during sleep. In: Sheldon SH, Ferber R, Kryger MH, Gozal D, editors. Principles and practice of pediatric sleep medicine 2nd ed. London: Elsevier; 2014. p. 179–194.

17. Hult A, Gjergja Juraški R, Gracia-Tabuenca J, Partinen M, Plavec D, Seppä V-P. Sources of variability in expiratory flow profiles during sleep in healthy young children. Respir Physiol Neurobiol 2019; 274: 103352.